\documentclass[aps,pra,twocolumn,superscriptaddress,amsfonts,floatfix,showpacs]{revtex4-1}
\usepackage{times}
\usepackage{ulem}
\usepackage{graphicx}
\usepackage{array}
\usepackage{amsmath}
\usepackage{color}
\begin{document}

\title{Simulating the discrete-time quantum walk dynamics with simultaneous coin and shift operators}

\author{J. Khatibi Moqadam}
\affiliation{Instituto de F\'\i sica ``Gleb Wataghin'', Universidade Estadual de Campinas (UNICAMP), Campinas, SP, Brazil}

\author{M. C. de Oliveira }
\affiliation{Instituto de F\'\i sica ``Gleb Wataghin'', Universidade Estadual de Campinas (UNICAMP), Campinas, SP, Brazil}


\date{\today}

\begin{abstract}
We implement the discrete-time quantum walk model using the continuous-time evolution of the Hamiltonian that includes both
the shift and the coin generators. Based on the Trotter-Suzuki first-order approximation, we consider an optimization problem in
which the Hellinger distance between the walker probability distributions resulted from the evolution of such Hamiltonian
and the quantum walk dynamics is minimized. The phase space implementation of the quantum walk is considered where the walker
state is encoded on the coherent state of a resonator and the coin on the two-level state of a qubit. In this approach, no
mechanism for switching between the coin and the shift operators is included. We show the Hellinger distance is bounded for
large number of time steps. The distance is small when we deviate from the standard quantum walk model, namely, when the
walker is allowed to move in between the sites. In simulating the standard quantum walk model, the distance is large but
bounded by $26\%$ for a relevant number of time steps. Even so, the system evolution shows the essential characteristics of
the standard quantum walk dynamics, namely, the ballistic evolution of the probability distribution and the linear growth of
the corresponding standard deviation. Moreover, the entanglement generated in this approach is approximately the same as the
entanglement generated in the standard quantum walk dynamics. Finally, the system dynamics under the influence of the
decoherence also shows the similar quantum-to-classical transition as in the standard quantum walk dynamics.
\end{abstract}
\maketitle

\section{Introduction}
Discrete-time quantum walks, the counterpart of the classical random walks, have recently attracted considerable attention
in designing quantum search algorithms that outperform the similar classical algorithms~\cite{Portugal2013}. It has been also
shown that the discrete-time quantum walks are universal for quantum computation~\cite{lovett2010universal}.
Many experimental setups, therefore, have been proposed to implement quantum walks and several implementations have also been
reported~\cite{Manouchehri2014}.
Examples of systems explored include micromasers~\cite{Aharonov1993},
ion traps~\cite{Travaglione2002,Schmitz2009,Zahringer2010}, cavity quantum electrodynamics~\cite{sanders2003quantum},
ensembles of nitrogen-vacancy centers in diamond~\cite{hardal2013discrete} and optomechanical
systems~\cite{moqadam2014quantum}.

In order to implement the discrete-time quantum walk model, two separate unitary operators are required, namely, the coin and
the shift operators. The corresponding Hamiltonians do not commute hence the quantum walk dynamics is generated by alternating
their actions. This is, however, technically difficult to implement depending on the physical system at hand.
The discrete-time quantum walks may also be realized by using an always-on shift Hamiltonian but applying a piecewise coin
Hamiltonian from time to time. A sequence of pulses is then required to implement the coin operator. The application of such
pulses, however, generally introduces noise and disturbs the walker's position.

In this way, the experimental realizations with the continuous-time quantum walk model~\cite{farhi1998quantum}, that involves
no coin operator, are usually more convenient and preferred. However, there is no known search algorithm based on the
continuous-time model that outperforms any classical algorithm, when the dimension of the data structure lattice is less than
four~\cite{childs2004spatial}. For the two- or three-dimensional lattices, efficient algorithms exist just whenever the lattice
spectrum possesses Dirac
points~\cite{childs2004spatial-dirac,foulger2014quantum,childs2014spatial-crystal,bohm2015microwave,ambainis2013search}. 
A strategy to overcome the problems of the implementation of the discrete-time quantum walk model, would be, therefore,
to keep the discrete-time quantum walk model but implementing the walk with the continuous-time evolution of the system.
The same strategy may be applied to coinless quantum walks, for example to the one described in~\cite{portugal2015one}.

In this paper, we employ the Trotter-Suzuki first-order approximation~\citep{suzuki1976generalized,suzuki1985decomposition},
exploring the situations in which the discrete-time quantum walk model can be efficiently simulated by the continuous-time
evolution of the total Hamiltonian that includes both the coin and the shift generators. In fact, the continuous-time evolution
of the total Hamiltonian when considered at some specified time steps approximately produces the discrete-time quantum
dynamics, with a variable amount of error. Having realized the coin and the shift operators in a single time step, we call
this approach the discrete-time quantum walk with simultaneous coin and shift operators.
In such implementation, the shift and the coin generators in the Hamiltonian are always on. Even so, the effects of the non
commutativity of those generators does not destroy the important features of the discrete-time quantum walk
dynamics. Those features include the ballistic evolution of the walker probability distribution, the linear growth of the
corresponding standard deviation, the amount of the generated coin-walker entanglement and the quantum-to-classical transition
of the probability distribution in the presence of the decoherence. The significance of such implementation in designing
efficient quantum algorithms, however, needs to be addressed elsewhere.

We specifically consider the quantum walk on a circle in the phase space with a two-sided coin, given its simplicity and
potentiality for implementation on a variety of physical systems \cite{Aharonov1993,Travaglione2002,Schmitz2009,Zahringer2010,
sanders2003quantum,hardal2013discrete,moqadam2014quantum}, but our findings extends to other geometries as well.
After a brief review of the standard implementation of the discrete-time quantum walk on a circle in Sec. \ref{2}, we show
the total Hamiltonian including both the shift and the coin parts can approximately simulate the quantum walk dynamics up to
a certain fidelity in Sec. \ref{3}. By modifying the Trotter-Suzuki approximation we optimize the system frequencies to minimize
the approximation error. Based on the walker jump at each time step we introduce a parameter that specifies the level of the
error.  We finally consider more closely other characteristics of both the continuous-time evolution and the standard
discrete-time quantum walk dynamics, such as the standard deviation, the coin-walker entanglement, as well as the effect of a
decoherent coin in Sec. \ref{4}. Finally in Sec. \ref{5} we present our conclusions.

\section{The standard quantum walk on a circle}\label{2}
The dynamics of the discrete-time quantum walk is described by
\begin{equation}
\label{eq:master_equation}
| \psi(t+1) \rangle = \mathcal{S}( \mathcal{I} \otimes \mathcal{C} ) | \psi(t) \rangle,
\end{equation}
where $\mathcal{S}$ and $\mathcal{C}$ are the shift and the coin operators, respectively, and $\mathcal{I}$ is the walker
space identity. At each time step, the state of the quantum walk $|\psi(t)\rangle$ is a superposition of the walker-coin
compound states. For a walk on a circle with $d$ sites the generic state of the quantum walk at time $t$ is given
by
\begin{equation}
\label{eq:generic_state}
| \psi(t) \rangle = \sum_{m=0}^{d-1} \bigl[\; \psi_0(m,t) |0 \rangle +
                                              \psi_1(m,t) |1 \rangle \;\bigr]\; | \varphi_m \rangle,
\end{equation}
where $\psi_{0,1}$ are the probability amplitudes in terms of the site
basis $\bigl\{| \varphi_m = 2 \pi m / d \rangle , m = 0 \ldots d-1 \bigr\}$
and the coin basis $\bigl\{| 0 \rangle , | 1 \rangle \bigr\}$.
The interesting quantity here is the walker probability distribution over the sites that is obtained by tracing out the coin
space. Unlike the behavior of the classical random walk, the dynamics of the probability distribution here is ballistic and
the corresponding standard deviation grows linearly with the time steps~\cite{Portugal2013}.

In the phase space realization of the quantum walk, the walker is encoded on the coherent state of a resonator field and
the coin on the two-level state of a qubit which is coupled to the resonator. The required Hamiltonian is
\begin{equation}
\label{eq:system_hamiltonian}
\mathcal{H} = \frac{1}{2} \hbar \omega \sigma_x - \hbar g a^{\dagger} a \sigma_z,
\end{equation}
where $\omega$ is the qubit transition frequency, $g$ is the qubit-resonator 
coupling constant,
$a^{\dagger}$ ($a$) is the resonator 
creation (annihilation) operators and $\sigma_x$ and $\sigma_z$
are the Pauli matrices. 

Suppose the initial state of the system is given by
$| \psi_0 \rangle = \nobreak | \alpha \rangle | q \rangle$ where $| \alpha \rangle$
is the coherent state of the resonator and $|q \rangle$ is the state
of the qubit. 
Suppose, furthermore, there is a mechanism that allows to switch between the first (the coin) and the second (the shift) terms in
Hamiltonian~(\ref{eq:system_hamiltonian}), alternately, during the evolution time $t$.
The challenge in the experimental realizations of the discrete-time quantum walk model is, actually, to effectively establish
such an ``alternating'' Hamiltonian. Nevertheless, consider a uniform time stepping of duration $t/n = \tau$ with which the
Hamiltonian is switched. By choosing the resonator frequency such that $\omega \tau = \pi/2$, the first term in
Hamiltonian~(\ref{eq:system_hamiltonian}) generates a Hadamard-like transformation
($\pi/2$-rotation around the $x$ axis) for the qubit, thus the coin
operator $\mathcal{C}$ is implemented. The second term generates a rotation for the coherent state in the phase space by the
angle $\pm g \tau$ whose sign depends on the state of the qubit. In this way, the shift operator $\mathcal{S}$ is realized.
Therefore, after each period $2\tau$ a single step of the quantum walk has been completed.

To explore the system dynamics, a finite dimensional phase space of size $d$ is required which is obtained by
truncating the infinite dimensional Fock space. The number operator $a^{\dagger} a$ is expected to be conjugate to the
phase operator. Therefore, the basis for the finite-dimensional phase space is constructed by calculating the discrete Fourier
transform of the number states in the truncated Fock space. The initial coefficients $\psi_{0,1}(m,0)$ in
Eq.~(\ref{eq:generic_state}) are determined by calculating the discrete Fourier transform of the initial truncated
coherent state and knowing the initial coin state. Note that the initial coherent state is a non-local state in the phase space
with a distribution close to the Gaussian.
The dynamics of the system with the ``alternating'' Hamiltonian is then obtained, using Eq.~(\ref{eq:master_equation}).
The shift operator in terms of the phase space basis is given by $\mathcal{F} \mathcal{S} \mathcal{F}^{-1}$
where $\mathcal{F}$ is the Fourier transform operator.
Choosing the coupling constant such that $g \tau = 2\pi/d$ forces the walker to hop between the sites on the circle.
The phase probability distribution of the walker, at any time step, is obtained by considering the diagonal elements of the
density operator of the system after tracing out the coin space. Experimentally, optical homodyne tomography can be
applied to reconstruct the state of the walker, if it is associated to an optical field, and ultimately used to reconstruct
the phase probability distribution~\citep{lvovsky2009continuous}.

The truncated dimension $d$ which is also the total number of sites for walking on the circle is highly restricted in
Ref.~\cite{sanders2003quantum}. There, it is demanded the sites are well separated in order to have approximately complete
overlap between the initial coherent state and a single phase state, hence to reach a local initial walker state.
Therefor, by fitting $2\pi |\alpha|$ distinguishable coherent state on a circle of radius $|\alpha|$,
the strict range $|\alpha|^2+|\alpha| \leq d \leq 2\pi |\alpha|$ for the truncated dimension is established. The lower bound
is required for the coherent state has reasonable support on the truncated Fock (or phase) space~\cite{sanders2003quantum}. Such
restrictions leads to the maximum number of sites $d_{\max}=33$ for $|\alpha|_{\max}=2\pi-1$.
However, treating the coherent state as a non-local initial state in the phase space makes the upper bound for the
dimension $d$ unnecessary, as far as the quantum walk dynamics is concerned.
But, if just the distinguishability of the coherent states on the circle is required a wider interval can be
considered. Actually, approximately $4\pi |\alpha|$ distinguishable coherent state can be fitted on
a circle with radius $|\alpha|$~\cite{moqadam2014quantum}. That can be justified by considering the geometrically obtained
uncertainty relation $\Delta \varphi \Delta n \geq 0.506$ for $|\alpha| \geq 1$. Here, $\Delta \varphi$ is the phase uncertainty
and $\Delta n = |\alpha|$ is the field excitation number uncertainty. In this paper, we consider the wider interval
which leads to the maximum number of sites $d_{\max}=145$ for $|\alpha|_{\max}=4\pi-1$.
It should be mentioned that for a given $|\alpha|$ choosing the largest allowed value of $d$ does not always lead to
the best quantum walk dynamics. Such cases, however, can be cured by changing the dimension.

\section{Simulation with simultaneous coin and shift operators}\label{3}
The above scenario is considered as the ideal way for implementing the discrete-time quantum
walk model. The basic problem in implementing such method is the technical difficulties in realizing a mechanism for switching
between the shift and the coin terms in Hamiltonian~(\ref{eq:system_hamiltonian}). In this section, we analyze the
simulation of the discrete-time quantum walk dynamics using simultaneous coin and shift operators. That can be realized by
continuous evolution of the system without needing any switching mechanism.

The time evolution of Hamiltonian~(\ref{eq:system_hamiltonian}) can be decomposed using
the Trotter-Suzuki approximation~\cite{suzuki1976generalized,suzuki1985decomposition}
\begin{equation}
\label{eq:trotter_suzuki}
\mathcal{U} = e^{ -i\mathcal{H}t/\hbar } \approx \left( e^{  i g a^{\dagger} a \sigma_z t/n }
                                                     \; e^{ -i \omega \sigma_x t/2n } \right)^n,
\end{equation}
where the approximation error scales as $g\omega t^2/n$ that can be made arbitrarily small by
increasing $n$.
The approximated evolution in Eq.~(\ref{eq:trotter_suzuki}) corresponds to a $n$-step discrete-time quantum walk
dynamics. At each time step $t/n=\tau$, the state of the qubit is transformed by the
rotation $\mathcal{R}_x(\omega \tau)$ about  the $x$ axis, hence the coin operator is realized. Depending on the state of the
qubit the coherent state rotates by the angle $\pm g \tau$ accordingly, thus the shift operator is also realized
(see~\cite{moqadam2014quantum}).
In this way, the operator $\mathcal{U}$ in Eq.~(\ref{eq:trotter_suzuki}) which is corresponding to the continuous-time evolution
of the system seems to be useful for simulating the discrete-time quantum walk dynamics.
Actually, the continuous-time evolution considered at time steps $\tau$ mimics a discrete-time quantum walk dynamics.
Note that while in the case of ``alternating'' Hamiltonian a single step of the quantum walk is completed in period $2\tau$,
in the case of Trotterized Hamiltonian it is completed in $\tau$.
Therefore, associating the latter case with the quantum walk with simultaneous coin and shift operators seems reasonable.

By setting the system frequencies and the time step such that $g\tau=2\pi/d$ and $\omega\tau=\pi/2$, the approximated evolution
in Eq.~(\ref{eq:trotter_suzuki}) generates a quantum walk dynamics with the Hadamard-like coin. Such dynamics, however, does not
correspond to the exact evolution of the system in Eq.~(\ref{eq:trotter_suzuki}). In fact with the mentioned setup,
the walker phase probability distributions resulted from the exact and the approximated evolutions will be very different after
a few time steps. We measure the difference between the probability distributions by the Hellinger
distance~\cite{pollard2002user}
\begin{equation}
\label{eq:Hellinger}
H(P_\mathrm{ex},P_\mathrm{app}) =
\frac{1}{\sqrt{2}} \| \sqrt{P_\mathrm{ex}}-\sqrt{P_\mathrm{app}} \|_{2},
\end{equation}
where $P_\mathrm{ex}$ and $P_\mathrm{app}$ are the walker phase probability distributions at a given step of the evolution,
corresponding to the exact and the approximate dynamics, respectively. The norm $\|.\|_2$ is the Euclidean norm.
We associate the distance in Eq.~(\ref{eq:Hellinger}) with the approximation error in Eq.~(\ref{eq:trotter_suzuki}). 

The approximation error in Eq.~(\ref{eq:trotter_suzuki}), can be decreased by decreasing the time step $\tau$.
But, decreasing $\tau$  will specifically decrease the angular rotation of the qubit $\omega \tau$, hence a coin operator
with $\omega\tau<\pi/2$ is realized. Specially, when a very small upper bound error is allowed, $\tau$ is required to be very
small and therefore the coin gets very close to the identity. With such a coin, the dynamics is almost deterministic and the
initial probability distribution is just displaced at each time step without deformation.

It is possible to decrease the error and still generate the optimal desired Hadamard coin
by decreasing $g$ but keeping $\omega$ and $\tau$ fixed. The error, however, will not go to zero just by
decreasing $g$. Very small $g$'s, make the dynamics very slow without really improving the approximation when the total
evolution time is of order $\tau$. This is the time order within which the ballistic spread of the probability distribution
is occurred.
In fact, the condition $\omega\tau=\pi/2$ for obtaining the Hadamard coin causes the Trotter-Suzuki approximation no longer
produces accurate results in time scales of $\tau$. The reason can be explained by considering the Zassenhaus expansion of the
system dynamics~\cite{suzuki1977convergence}. In the Zassenhaus expansion of the exact evolution in Eq.~(\ref{eq:trotter_suzuki}),
which is actually an infinite product of exponential operators, exponents like $(\omega\tau)^k(g\tau)^l$ appears where $k$
and $l$ are positive integer numbers. Therefore, with a Hadamard coin, it is not possible to obtain a good approximation by
keeping just the first two exponentials in the expansion no matter how small $g$ is.

However, in the above case where $g$ is decreased and the other parameters are kept fixed, the difference between the phase
probability distributions resulted from the exact and the approximated dynamics in Eq.~(\ref{eq:trotter_suzuki}) seems to be
mainly due to the displacement of the distributions.
To improve the approximation we therefore slightly modify the Trotter-Suzuki approximation in a single time step
\begin{equation}
\label{eq:ts_imp}
e^{ i ( g \tau a^{\dagger}a \sigma_z \;-\; \omega \tau \sigma_x / 2 ) } \approx
                       e^{ i (2\pi/rd) a^{\dagger}a \sigma_z } \; e^{ -i (\pi/4) \sigma_x},
\end{equation}
where the dimensionless frequencies $g\tau$ and $\omega\tau$ need to be optimized such that the approximation error is minimum.
The parameter $r$ is a positive number that is used for controlling the angular step of the simulated quantum walk.
The idea here is to adjust the system frequencies such that the continuous-time evolution of the system coincide the desired
discrete-time quantum walk dynamics at time steps $\tau$, as mush as possible. The dimensionless frequencies
are optimized by minimizing the average Hellinger distance in Eq.~(\ref{eq:Hellinger}) calculated for large number of time steps.
To be close enough to the global optimal solution, the maximization process is carried over 100 random initial guesses for the
frequencies and then the solution with the smallest distance is selected.

\begin{figure}[t]
\includegraphics[trim = 11mm 16mm 0mm 0mm, clip=true, width=8.8cm]{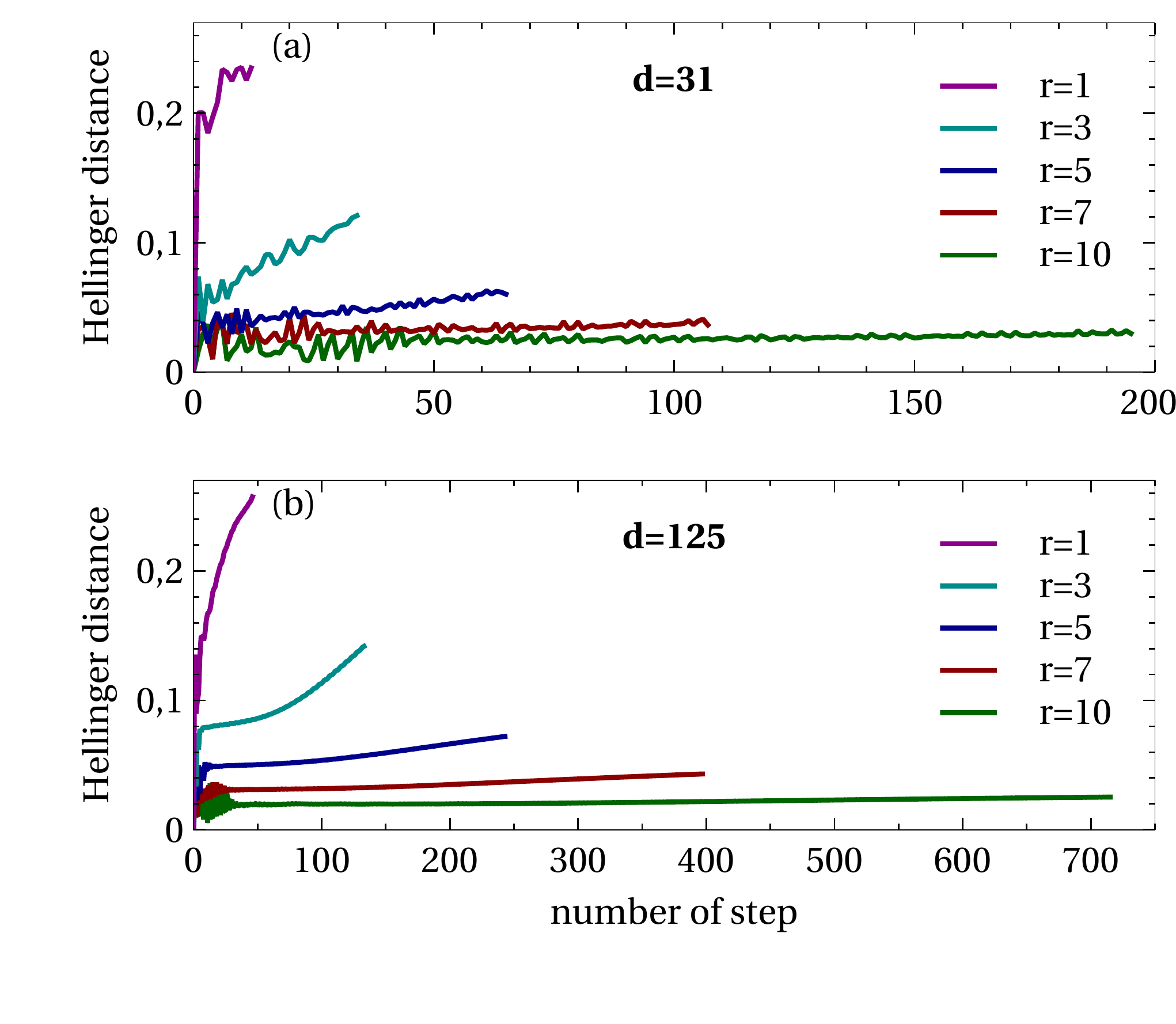}
\caption{(Color Online) The Hellinger distance between the two probability distributions resulted from the exact and the
approximated evolutions in Eq.~(\ref{eq:ts_imp}), in terms of the time steps. The distances are given for five values of $r$
in the case $d=31$ and $\alpha = -5$ in panel (a) and the case $d=125$ and $\alpha = -10$ in panel (b). In both cases the
initial state of the qubit is set to the ground state.}
\label{fig:errors}
\end{figure}

Figure~\ref{fig:errors} shows the Hellinger distance between the phase probability distributions resulted from the exact and the
approximated evolutions in Eq.~(\ref{eq:ts_imp}), for different values of $r$. Two cases with the phase space dimensions $d=31,125$
and corresponding average field excitation numbers $|\alpha|=5,10$ have been considered here. In both cases the initial state of
the qubit was set to the ground state $|q\rangle=|0\rangle$. Moreover, the total number of steps at each $r$ is set such that
the initial distribution is split into two peaks with the angular distance of $\pi$ radian.
The diagrams show that by increasing $r$ the approximation error over the time steps decreases. Increasing $r$ can be
associated with decreasing $g$ while keeping fixed $\omega$ and $\tau$, that improves the approximation as
described before.
\begin{figure}[t]
\includegraphics[trim = 11mm 16mm 0mm 0mm, clip=true, width=8.8cm]{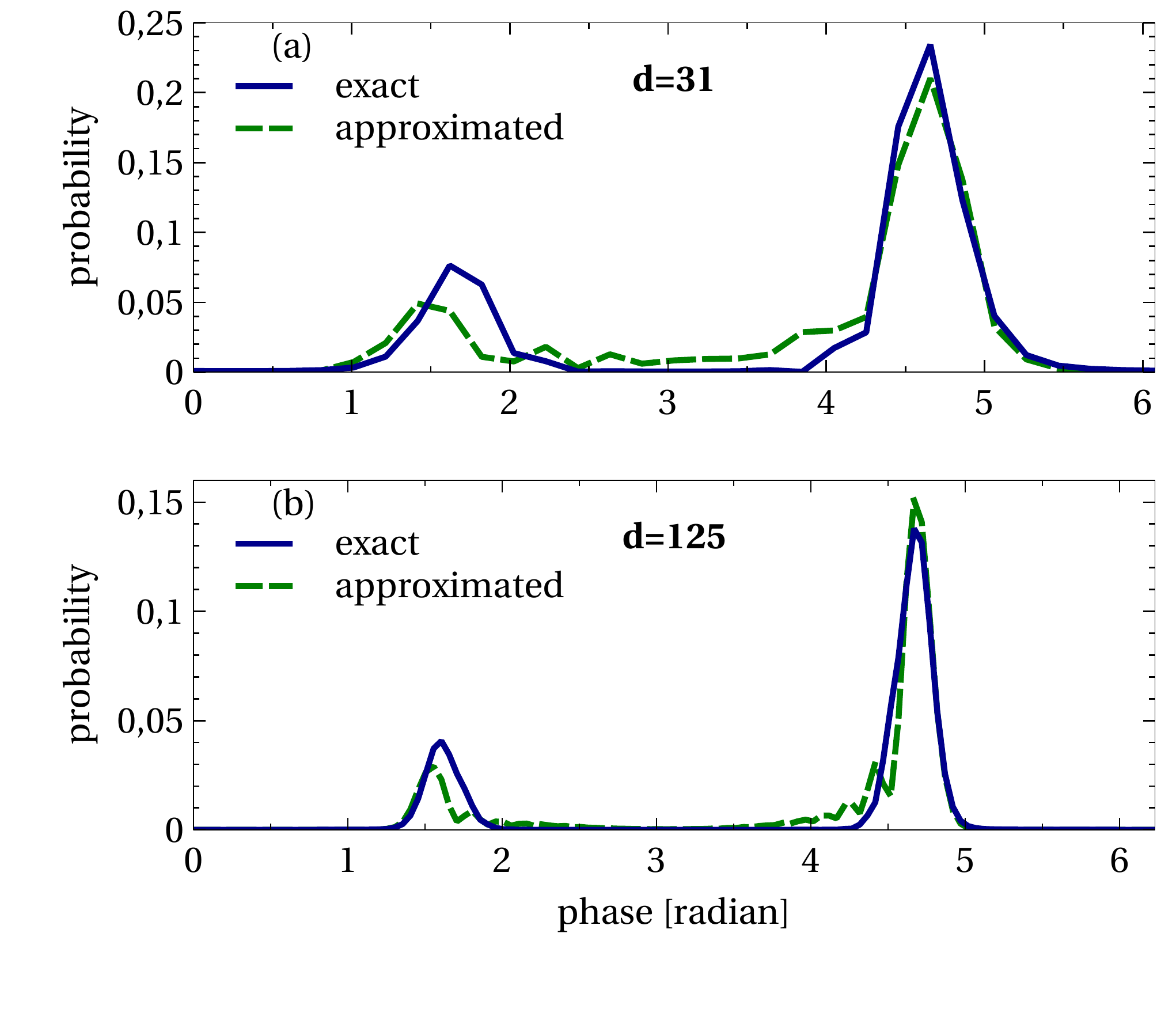}
\caption{(Color Online) The phase probability distributions for the exact (solid line) and the approximated
(dashed line) evolutions in Eq.~(\ref{eq:ts_imp}) with $r=1$, when the peaks are separated by $\pi$ radian.
Panels (a) and (b) correspond to the cases $d=31$ $(\alpha = -5)$ and $d=125$ $(\alpha = -10)$, respectively.
In both cases the initial state of the qubit is set to the ground state.}
\label{fig:r1_distributions}
\end{figure}
As previously mentioned, the phase space dimension $d$ corresponds to the total number of sites on the circle and specifies
the walker Hilbert space dimension. Comparing panels (a) and (b) in~Fig.~\ref{fig:errors}, shows that the effect of increasing
the dimension $d$ is just to diminish the oscillations in the error plots.

According to Fig.~\ref{fig:errors}, for the case $r=10$ (green/lowest plots in both panels) the approximation error is almost
stable and remains below $4\%$ for a large number of time steps. However, since the angular distance between the sites is $2\pi/d$,
choosing $r=10$ causes the walker is conditionally shifted by $1/10$ of the distance
between the sites. So, in spite of the good match between the exact and the approximated evolutions in this case,
the approximated evolution does not represent the standard quantum walk evolution. In the standard model the walker is
restricted to jump between the neighbor sites at each time step.
Nevertheless, the standard deviation of the phase probability distribution in both exact and approximated dynamics is still
linearly proportional to the number of steps but the corresponding slopes drop at least one order lower, comparing the
standard quantum walk case.

For the case $r=1$ (magenta/highest plots in both panels) in Fig.~\ref{fig:errors}, the approximated dynamics in
Eq.~(\ref{eq:ts_imp}) corresponds to the standard quantum walk evolution. In this case, however, the error is larger but
remains below $26\%$ until the initial distributions are split into two peaks with the angular distance of $\pi$ radian.
To get an idea of how the two dynamics are different at such error the corresponding phase probability distributions are
presented in the next figure.

Figure~\ref{fig:r1_distributions} shows the phase probability distributions for the exact (solid line) and the approximated
(dashed line) evolutions corresponding to the case $r=1$ in Eq.~(\ref{eq:ts_imp}), for two different
dimensions $d=\nobreak 31,125$. The plots show the probability distributions for the time step when the two peaks are
separated by the angular distance of $\pi$ radian. It can be seen that,
the probability distributions are qualitatively similar, representing two major peaks.
The exact dynamics, therefore, represents one of the important features of the quantum walks, the ballistic spread of the walker
probability distribution. A special difference between the two plots corresponds to the intermediate interference between
the two peaks related to the approximated evolution (the standard quantum walk dynamics). For the exact evolution
the probability distribution is almost flat between the two peaks.

Actually, it is not possible to reproduce the intermediate interference with just playing with the system frequencies.
It is, nevertheless, tempting to think of including more terms in the Hamiltonian of the exact evolution in
Eq.~(\ref{eq:ts_imp}) to improve the approximation. Specially, higher order terms from the Baker-Campbell-Hausdorff (BCH)
formula~\cite{suzuki1977convergence} may seem suitable in this respect.
However, according to the convergence theorem proved in Ref.~\cite{suzuki1977convergence}
the BCH formula does not converge because of the large argument $\omega\tau=\pi/2$ of the Hadamard coin.
Therefore, the higher order terms from BCH formula does not lead to an improved approximation.
Even so, we have numerically verified that adding a term proportional to $a^{\dagger}a \sigma_x$ can improve the
approximation to some degree. It may be stressed that this term is different from the second order term that comes from
the BCH formula and is proportional to $a^{\dagger}a \sigma_y$. We are not going to consider such term in this paper.

\begin{figure*}
\includegraphics[trim = 15mm 19mm 0mm 0mm, clip=true, width=16cm]{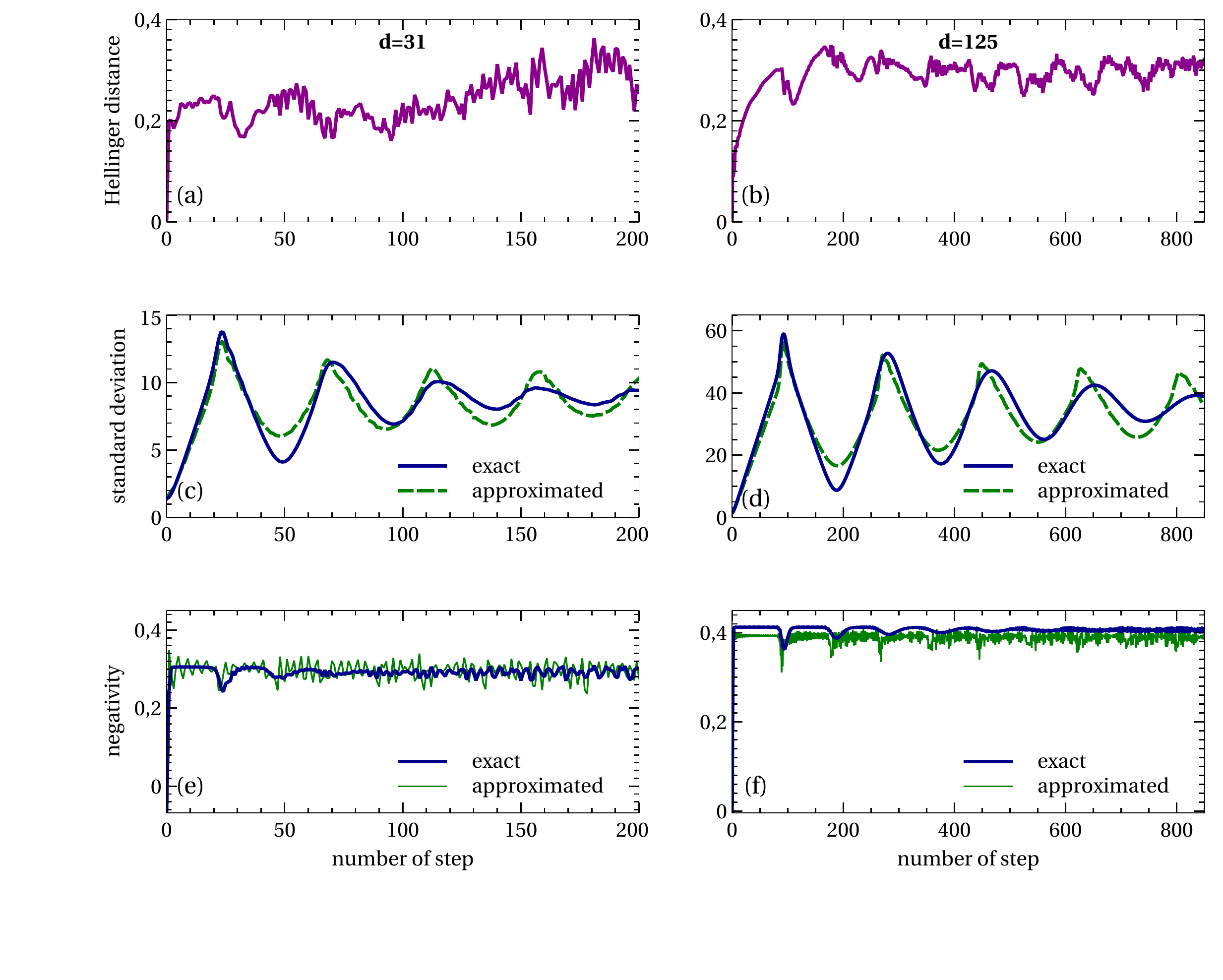}
\caption{(Color Online) Different characteristics of the exact and the approximated
evolutions in Eq.~(\ref{eq:ts_imp}), for the case $r=1$. The left and the right panels correspond to the cases
$d=31$ $(\alpha = -5)$ and $d=125$ $(\alpha = -10)$, respectively. Panels (a) and (b) show the Hellinger distance between the
corresponding probability distributions for large number of time steps. Panels (c) and (d) show the standard deviations of the
probability distributions over the time steps for the exact (solid line) and the approximated (dashed line) evolutions.
The negativity entanglement between the qubit and the resonator for the exact/approximated evolutions is showed
by the thick/thin line in panels (e) and (f).}
\label{fig:distributions}
\end{figure*}

We now investigate in more details the case $r=1$ in Eq.~(\ref{eq:ts_imp}), which corresponds to the standard quantum
walk dynamics. Panels (a) and (b) in Fig.~\ref{fig:distributions} show the Hellinger distance between the exact and the
approximated probability distributions for large number of time steps, in the two cases $d=31,125$.
The plots in this panels are actually the continuations of the plots in Fig.~\ref{fig:errors}, for the case $r=1$.
The system frequencies for producing all the plots have been optimized by minimizing the average Hellinger distance over the
first $12$ time steps for the dimension $31$, and the first $46$ steps for the dimension $125$. However, including more time
steps in the optimization process will lead to a more balanced error plots, specially in the case $d=31$ presented in panel (a).

For time steps less than the total number of sites on the circle, the exact dynamics shows no interference
(see Fig.~\ref{fig:r1_distributions}). However, the bounded error in Fig.~\ref{fig:distributions} indicates that as time step
goes beyond the total number of sites on the circle the interference is generated. We have verified (not showed here) the two
major peaks in Fig.~\ref{fig:r1_distributions} (solid lines) really interfere when pass through each other.

The standard deviations of the probability distributions related to Eq.~(\ref{eq:ts_imp}) are showed in panels (c) and
(d) in Fig.~\ref{fig:distributions}, for the two cases $d=31,125$. The solid/dashed line corresponds to the
exact/approximated evolution. After the initial distribution is split into two peaks with the angular distance around $2\pi$
radian, the peaks will move toward each other hence the standard deviation starts to decrease. The oscillations of the standard
deviation in the figure indicates the quantum walk spreads around the circle for several times.
There is a good match between the two plots at least for time steps less than the available sites on the circle.
The exact evolution, therefore represents another important feature of the quantum walks, the linear growth of the standard
deviation of the walker probability distribution.

Finally, the  entanglement between the qubit and the resonator is given by the negativity in panels (g) and (h)
in Fig.~\ref{fig:distributions}. The thick/thin line corresponds to the exact/approximated evolution in Eq.~(\ref{eq:ts_imp}).
The figure shows there is no important difference between the entanglement that is generated during the evolution in both the
exact and the approximated evolutions. The entanglement dynamics is just smoother for the exact evolution. 

Figures \ref{fig:r1_distributions} and \ref{fig:distributions}, therefore, suggest the discrete-time quantum walk with
simultaneous coin and shift operators is promising in simulating the standard discrete-time quantum walk model. Despite some
differences, the implementation with simultaneous coin and shift operators have the important features of the standard
quantum walk, namely, the ballistic spread of the walker probability distribution and the linear dependency of the corresponding
standard deviation on the time steps. Moreover, the amount of the generated entanglement between the coin and the walker is
almost the same level as the one generated in the standard quantum walk model.

Inspecting the optimized system frequencies in Eq.~(\ref{eq:ts_imp}) reveals that the ratio
\begin{equation}
C = \frac{ \omega\tau / g\tau } { \left( \pi/2 \right) / \left( 2\pi/rd \right) } = \frac{4\omega}{rdg}
\end{equation}
approximately amounts to $10$, in the case of $r=1$. As the optimized value of $g\tau$ is not very different
from $2\pi/d$ the optimized value of $\omega\tau$ is about $C\pi/2$. This is true for both dimensions
$d=31,125$ as well as the other relevant dimensions.
Therefore, the system frequencies can be adjusted very easily to obtain an approximate quantum walk dynamics with the exact
evolution in Eq.~(\ref{eq:ts_imp}). We have verified numerically the ratio $C$ depends on the coin that is used in the quantum
walk dynamics, namely the qubit rotation $\mathcal{R}_x(\omega\tau)$ around the $x$ axis. The value of $C$ changes in the
direction of changing the angle of rotation. Decreasing the angle from $\pi/2$ (changing the coin toward the identity)
decreases $C$, and in the contrary, increasing the angle from $\pi/2$ (changing the coin toward the completely biased coin)
increases $C$. Actually, the different scales of the system frequencies allows to develop the desired qubit rotation.
In this way, it is possible to implement any desired coin.

\section{Decoherent coin}\label{4}
As mentioned before, the probability distribution of the quantum walk spreads ballistically and the corresponding standard
deviation evolves linearly in terms of the time steps. In the presence of the decoherence, however, the quantum walk makes a
transition to a random walk in which the spread of the probability distribution is Gaussian and the standard deviation
evolves with the square root of the time steps.

Simulating the effect of the dephasing channel~\cite{nielsen2010quantum} on the two-sided coin (the qubit) for both the exact
and the approximated dynamics in Eq.~(\ref{eq:ts_imp}) shows qualitatively the same behaviors for the case $r=1$. For a given
dephasing channel, the Hellinger distance between the corresponding probability distributions decreases when $r$ increases.
Moreover, the Hellinger distance is decreasing in terms of the time steps for nonzero decoherence. Actually, in the
classical limit both the exact and the approximated dynamics become identical.

\section{Conclusions}\label{5}
In this paper, we have analyzed the simulation of the discrete-time quantum walk dynamics, using the continuous-time
evolution of a coupled resonator-qubit system. The coherent state of the resonator is served as the walker and the
state of the qubit as the two-sided coin. Instead of successive applications of the coin and the shift operators, we apply
a single operator which is generated by the evolution of the Hamiltonian that includes both the coin
and the shift generators. Actually, the continuous-time evolution of the system considered at the specified time steps imitate
the discrete-time quantum walk dynamics. We have called this method the discrete-time quantum walk with simultaneous coin and
shift operators. In this approach, no mechanism for switching between the coin and the shift operators is included. In fact,
the corresponding operators are implemented in a common single time step.

Based on the Trotter-Suzuki approximation, we wrote an optimization problem whose global solutions (the system frequencies)
made the approximation error minimum. The approximation error is calculated by the Hellinger distance between the walker
probability distributions resulted from the system dynamics and the quantum walk dynamics. We showed the error is bounded
for large number of time steps. The error is small when we deviate from the standard quantum walk model, namely, when the walker
is allowed to move in between the sites. For the standard quantum walk model, the error is large and bounded by $26\%$
for a relevant number of time steps.

In spite of that error, the system evolution with simultaneous coin and shift operators shows the essential characteristics of
the standard quantum walk dynamics. The probability distribution spreads ballistically and the corresponding standard deviation
grows linearly in time. The entanglement generated in this method is also approximately the same as the entanglement generated in
the standard quantum walk dynamics. Moreover, the system dynamics under the influence of the decoherence shows the similar
quantum-to-classical transition as in the standard quantum walk dynamics.

It is expected the discrete-time quantum walk model with simultaneous coin and shift operators to be useful in implementing
quantum algorithms that outperforms classical ones. It has been proved the discrete-time quantum walks are universal for quantum
computing~\cite{lovett2010universal}. Actually, the quantum circuit model can be transformed into a graph on which the quantum
walk propagates and implements the quantum algorithm. We expect the similar property for the quantum walk with simultaneous coin
and shift operators.
However, at least a two-dimensional quantum walk is required to implement universal quantum computation. The extension of the
above model to more than one dimension is feasible, in principle. It is therefore interesting to check in what sense would the
above model be universal for quantum computing.

\begin{acknowledgments}
We  thank R. Portugal who raised the initial question and motivates us to explore the problem.
JKM acknowledges financial support from Brazilian National Council for Scientific and Technological Development
(CNPq), grant PDJ 165941/2014-6. MCO acknowledges support by FAPESP and CNPq through the National Institute for Science and
Technology of Quantum Information (INCT-IQ) and the Research Center in Optics and Photonics (CePOF).
\end{acknowledgments}

\bibliography{zbib}

\begin{thebibliography}{24}%
\makeatletter
\providecommand \@ifxundefined [1]{%
 \@ifx{#1\undefined}
}%
\providecommand \@ifnum [1]{%
 \ifnum #1\expandafter \@firstoftwo
 \else \expandafter \@secondoftwo
 \fi
}%
\providecommand \@ifx [1]{%
 \ifx #1\expandafter \@firstoftwo
 \else \expandafter \@secondoftwo
 \fi
}%
\providecommand \natexlab [1]{#1}%
\providecommand \enquote  [1]{``#1''}%
\providecommand \bibnamefont  [1]{#1}%
\providecommand \bibfnamefont [1]{#1}%
\providecommand \citenamefont [1]{#1}%
\providecommand \href@noop [0]{\@secondoftwo}%
\providecommand \href [0]{\begingroup \@sanitize@url \@href}%
\providecommand \@href[1]{\@@startlink{#1}\@@href}%
\providecommand \@@href[1]{\endgroup#1\@@endlink}%
\providecommand \@sanitize@url [0]{\catcode `\\12\catcode `\$12\catcode
  `\&12\catcode `\#12\catcode `\^12\catcode `\_12\catcode `\%12\relax}%
\providecommand \@@startlink[1]{}%
\providecommand \@@endlink[0]{}%
\providecommand \url  [0]{\begingroup\@sanitize@url \@url }%
\providecommand \@url [1]{\endgroup\@href {#1}{\urlprefix }}%
\providecommand \urlprefix  [0]{URL }%
\providecommand \Eprint [0]{\href }%
\providecommand \doibase [0]{http://dx.doi.org/}%
\providecommand \selectlanguage [0]{\@gobble}%
\providecommand \bibinfo  [0]{\@secondoftwo}%
\providecommand \bibfield  [0]{\@secondoftwo}%
\providecommand \translation [1]{[#1]}%
\providecommand \BibitemOpen [0]{}%
\providecommand \bibitemStop [0]{}%
\providecommand \bibitemNoStop [0]{.\EOS\space}%
\providecommand \EOS [0]{\spacefactor3000\relax}%
\providecommand \BibitemShut  [1]{\csname bibitem#1\endcsname}%
\let\auto@bib@innerbib\@empty
\bibitem [{\citenamefont {Portugal}(2013)}]{Portugal2013}%
  \BibitemOpen
  \bibfield  {author} {\bibinfo {author} {\bibfnamefont {R.}~\bibnamefont
  {Portugal}},\ }\href@noop {} {\emph {\bibinfo {title} {{Quantum Walks and
  Search Algorithms}}}}\ (\bibinfo  {publisher} {Springer},\ \bibinfo {year}
  {2013})\BibitemShut {NoStop}%
\bibitem [{\citenamefont {Lovett}\ \emph {et~al.}(2010)\citenamefont {Lovett},
  \citenamefont {Cooper}, \citenamefont {Everitt}, \citenamefont {Trevers},\
  and\ \citenamefont {Kendon}}]{lovett2010universal}%
  \BibitemOpen
  \bibfield  {author} {\bibinfo {author} {\bibfnamefont {N.~B.}\ \bibnamefont
  {Lovett}}, \bibinfo {author} {\bibfnamefont {S.}~\bibnamefont {Cooper}},
  \bibinfo {author} {\bibfnamefont {M.}~\bibnamefont {Everitt}}, \bibinfo
  {author} {\bibfnamefont {M.}~\bibnamefont {Trevers}}, \ and\ \bibinfo
  {author} {\bibfnamefont {V.}~\bibnamefont {Kendon}},\ }\href {\doibase
  10.1103/PhysRevA.81.042330} {\bibfield  {journal} {\bibinfo  {journal} {Phys.
  Rev. A}\ }\textbf {\bibinfo {volume} {81}},\ \bibinfo {pages} {042330}
  (\bibinfo {year} {2010})}\BibitemShut {NoStop}%
\bibitem [{\citenamefont {Manouchehri}\ and\ \citenamefont
  {Wang}(2014)}]{Manouchehri2014}%
  \BibitemOpen
  \bibfield  {author} {\bibinfo {author} {\bibfnamefont {K.}~\bibnamefont
  {Manouchehri}}\ and\ \bibinfo {author} {\bibfnamefont {J.}~\bibnamefont
  {Wang}},\ }\href@noop {} {\emph {\bibinfo {title} {{Physical Implementation
  of Quantum Walks}}}}\ (\bibinfo  {publisher} {Springer},\ \bibinfo {address}
  {Berlin, Heidelberg},\ \bibinfo {year} {2014})\BibitemShut {NoStop}%
\bibitem [{\citenamefont {Aharonov}\ \emph {et~al.}(1993)\citenamefont
  {Aharonov}, \citenamefont {Davidovich},\ and\ \citenamefont
  {Zagury}}]{Aharonov1993}%
  \BibitemOpen
  \bibfield  {author} {\bibinfo {author} {\bibfnamefont {Y.}~\bibnamefont
  {Aharonov}}, \bibinfo {author} {\bibfnamefont {L.}~\bibnamefont
  {Davidovich}}, \ and\ \bibinfo {author} {\bibfnamefont {N.}~\bibnamefont
  {Zagury}},\ }\href@noop {} {\bibfield  {journal} {\bibinfo  {journal}
  {Physical Review A}\ }\textbf {\bibinfo {volume} {48}},\ \bibinfo {pages}
  {1687} (\bibinfo {year} {1993})}\BibitemShut {NoStop}%
\bibitem [{\citenamefont {Travaglione}\ and\ \citenamefont
  {Milburn}(2002)}]{Travaglione2002}%
  \BibitemOpen
  \bibfield  {author} {\bibinfo {author} {\bibfnamefont {B.}~\bibnamefont
  {Travaglione}}\ and\ \bibinfo {author} {\bibfnamefont {G.}~\bibnamefont
  {Milburn}},\ }\href@noop {} {\bibfield  {journal} {\bibinfo  {journal}
  {Physical Review A}\ }\textbf {\bibinfo {volume} {65}},\ \bibinfo {pages}
  {032310} (\bibinfo {year} {2002})}\BibitemShut {NoStop}%
\bibitem [{\citenamefont {Schmitz}\ \emph {et~al.}(2009)\citenamefont
  {Schmitz}, \citenamefont {Matjeschk}, \citenamefont {Schneider},
  \citenamefont {Glueckert}, \citenamefont {Enderlein}, \citenamefont {Huber},\
  and\ \citenamefont {Schaetz}}]{Schmitz2009}%
  \BibitemOpen
  \bibfield  {author} {\bibinfo {author} {\bibfnamefont {H.}~\bibnamefont
  {Schmitz}}, \bibinfo {author} {\bibfnamefont {R.}~\bibnamefont {Matjeschk}},
  \bibinfo {author} {\bibfnamefont {C.}~\bibnamefont {Schneider}}, \bibinfo
  {author} {\bibfnamefont {J.}~\bibnamefont {Glueckert}}, \bibinfo {author}
  {\bibfnamefont {M.}~\bibnamefont {Enderlein}}, \bibinfo {author}
  {\bibfnamefont {T.}~\bibnamefont {Huber}}, \ and\ \bibinfo {author}
  {\bibfnamefont {T.}~\bibnamefont {Schaetz}},\ }\href@noop {} {\bibfield
  {journal} {\bibinfo  {journal} {Physical Review Letters}\ }\textbf {\bibinfo
  {volume} {103}},\ \bibinfo {pages} {090504} (\bibinfo {year}
  {2009})}\BibitemShut {NoStop}%
\bibitem [{\citenamefont {Z{\"a}hringer}\ \emph {et~al.}(2010)\citenamefont
  {Z{\"a}hringer}, \citenamefont {Kirchmair}, \citenamefont {Gerritsma},
  \citenamefont {Solano}, \citenamefont {Blatt},\ and\ \citenamefont
  {Roos}}]{Zahringer2010}%
  \BibitemOpen
  \bibfield  {author} {\bibinfo {author} {\bibfnamefont {F.}~\bibnamefont
  {Z{\"a}hringer}}, \bibinfo {author} {\bibfnamefont {G.}~\bibnamefont
  {Kirchmair}}, \bibinfo {author} {\bibfnamefont {R.}~\bibnamefont
  {Gerritsma}}, \bibinfo {author} {\bibfnamefont {E.}~\bibnamefont {Solano}},
  \bibinfo {author} {\bibfnamefont {R.}~\bibnamefont {Blatt}}, \ and\ \bibinfo
  {author} {\bibfnamefont {C.~F.}\ \bibnamefont {Roos}},\ }\href@noop {}
  {\bibfield  {journal} {\bibinfo  {journal} {Physical Review Letters}\
  }\textbf {\bibinfo {volume} {104}},\ \bibinfo {pages} {100503} (\bibinfo
  {year} {2010})}\BibitemShut {NoStop}%
\bibitem [{\citenamefont {Sanders}\ \emph {et~al.}(2003)\citenamefont
  {Sanders}, \citenamefont {Bartlett}, \citenamefont {Tregenna},\ and\
  \citenamefont {Knight}}]{sanders2003quantum}%
  \BibitemOpen
  \bibfield  {author} {\bibinfo {author} {\bibfnamefont {B.~C.}\ \bibnamefont
  {Sanders}}, \bibinfo {author} {\bibfnamefont {S.~D.}\ \bibnamefont
  {Bartlett}}, \bibinfo {author} {\bibfnamefont {B.}~\bibnamefont {Tregenna}},
  \ and\ \bibinfo {author} {\bibfnamefont {P.~L.}\ \bibnamefont {Knight}},\
  }\href@noop {} {\bibfield  {journal} {\bibinfo  {journal} {Physical Review
  A}\ }\textbf {\bibinfo {volume} {67}},\ \bibinfo {pages} {042305} (\bibinfo
  {year} {2003})}\BibitemShut {NoStop}%
\bibitem [{\citenamefont {Hardal}\ \emph {et~al.}(2013)\citenamefont {Hardal},
  \citenamefont {Xue}, \citenamefont {Shikano}, \citenamefont
  {M{\"u}stecapl{\i}o\u{g}lu},\ and\ \citenamefont
  {Sanders}}]{hardal2013discrete}%
  \BibitemOpen
  \bibfield  {author} {\bibinfo {author} {\bibfnamefont {A.~{\"U}.}\
  \bibnamefont {Hardal}}, \bibinfo {author} {\bibfnamefont {P.}~\bibnamefont
  {Xue}}, \bibinfo {author} {\bibfnamefont {Y.}~\bibnamefont {Shikano}},
  \bibinfo {author} {\bibfnamefont {{\"O}.~E.}\ \bibnamefont
  {M{\"u}stecapl{\i}o\u{g}lu}}, \ and\ \bibinfo {author} {\bibfnamefont
  {B.~C.}\ \bibnamefont {Sanders}},\ }\href@noop {} {\bibfield  {journal}
  {\bibinfo  {journal} {Physical Review A}\ }\textbf {\bibinfo {volume} {88}},\
  \bibinfo {pages} {022303} (\bibinfo {year} {2013})}\BibitemShut {NoStop}%
\bibitem [{\citenamefont {Moqadam}\ \emph {et~al.}(2014)\citenamefont
  {Moqadam}, \citenamefont {Portugal},\ and\ \citenamefont
  {de~Oliveira}}]{moqadam2014quantum}%
  \BibitemOpen
  \bibfield  {author} {\bibinfo {author} {\bibfnamefont {J.~K.}\ \bibnamefont
  {Moqadam}}, \bibinfo {author} {\bibfnamefont {R.}~\bibnamefont {Portugal}}, \
  and\ \bibinfo {author} {\bibfnamefont {M.~C.}\ \bibnamefont {de~Oliveira}},\
  }\href@noop {} {\bibfield  {journal} {\bibinfo  {journal} {arXiv preprint
  arXiv:1403.5205}\ } (\bibinfo {year} {2014})}\BibitemShut {NoStop}%
\bibitem [{\citenamefont {Farhi}\ and\ \citenamefont
  {Gutmann}(1998)}]{farhi1998quantum}%
  \BibitemOpen
  \bibfield  {author} {\bibinfo {author} {\bibfnamefont {E.}~\bibnamefont
  {Farhi}}\ and\ \bibinfo {author} {\bibfnamefont {S.}~\bibnamefont
  {Gutmann}},\ }\href {\doibase 10.1103/PhysRevA.58.915} {\bibfield  {journal}
  {\bibinfo  {journal} {Phys. Rev. A}\ }\textbf {\bibinfo {volume} {58}},\
  \bibinfo {pages} {915} (\bibinfo {year} {1998})}\BibitemShut {NoStop}%
\bibitem [{\citenamefont {Childs}\ and\ \citenamefont
  {Goldstone}(2004{\natexlab{a}})}]{childs2004spatial}%
  \BibitemOpen
  \bibfield  {author} {\bibinfo {author} {\bibfnamefont {A.~M.}\ \bibnamefont
  {Childs}}\ and\ \bibinfo {author} {\bibfnamefont {J.}~\bibnamefont
  {Goldstone}},\ }\href {\doibase 10.1103/PhysRevA.70.022314} {\bibfield
  {journal} {\bibinfo  {journal} {Phys. Rev. A}\ }\textbf {\bibinfo {volume}
  {70}},\ \bibinfo {pages} {022314} (\bibinfo {year}
  {2004}{\natexlab{a}})}\BibitemShut {NoStop}%
\bibitem [{\citenamefont {Childs}\ and\ \citenamefont
  {Goldstone}(2004{\natexlab{b}})}]{childs2004spatial-dirac}%
  \BibitemOpen
  \bibfield  {author} {\bibinfo {author} {\bibfnamefont {A.~M.}\ \bibnamefont
  {Childs}}\ and\ \bibinfo {author} {\bibfnamefont {J.}~\bibnamefont
  {Goldstone}},\ }\href {\doibase 10.1103/PhysRevA.70.042312} {\bibfield
  {journal} {\bibinfo  {journal} {Phys. Rev. A}\ }\textbf {\bibinfo {volume}
  {70}},\ \bibinfo {pages} {042312} (\bibinfo {year}
  {2004}{\natexlab{b}})}\BibitemShut {NoStop}%
\bibitem [{\citenamefont {Foulger}\ \emph {et~al.}(2014)\citenamefont
  {Foulger}, \citenamefont {Gnutzmann},\ and\ \citenamefont
  {Tanner}}]{foulger2014quantum}%
  \BibitemOpen
  \bibfield  {author} {\bibinfo {author} {\bibfnamefont {I.}~\bibnamefont
  {Foulger}}, \bibinfo {author} {\bibfnamefont {S.}~\bibnamefont {Gnutzmann}},
  \ and\ \bibinfo {author} {\bibfnamefont {G.}~\bibnamefont {Tanner}},\ }\href
  {\doibase 10.1103/PhysRevLett.112.070504} {\bibfield  {journal} {\bibinfo
  {journal} {Phys. Rev. Lett.}\ }\textbf {\bibinfo {volume} {112}},\ \bibinfo
  {pages} {070504} (\bibinfo {year} {2014})}\BibitemShut {NoStop}%
\bibitem [{\citenamefont {Childs}\ and\ \citenamefont
  {Ge}(2014)}]{childs2014spatial-crystal}%
  \BibitemOpen
  \bibfield  {author} {\bibinfo {author} {\bibfnamefont {A.~M.}\ \bibnamefont
  {Childs}}\ and\ \bibinfo {author} {\bibfnamefont {Y.}~\bibnamefont {Ge}},\
  }\href {\doibase 10.1103/PhysRevA.89.052337} {\bibfield  {journal} {\bibinfo
  {journal} {Phys. Rev. A}\ }\textbf {\bibinfo {volume} {89}},\ \bibinfo
  {pages} {052337} (\bibinfo {year} {2014})}\BibitemShut {NoStop}%
\bibitem [{\citenamefont {B\"ohm}\ \emph {et~al.}(2015)\citenamefont {B\"ohm},
  \citenamefont {Bellec}, \citenamefont {Mortessagne}, \citenamefont {Kuhl},
  \citenamefont {Barkhofen}, \citenamefont {Gehler}, \citenamefont
  {St\"ockmann}, \citenamefont {Foulger}, \citenamefont {Gnutzmann},\ and\
  \citenamefont {Tanner}}]{bohm2015microwave}%
  \BibitemOpen
  \bibfield  {author} {\bibinfo {author} {\bibfnamefont {J.}~\bibnamefont
  {B\"ohm}}, \bibinfo {author} {\bibfnamefont {M.}~\bibnamefont {Bellec}},
  \bibinfo {author} {\bibfnamefont {F.}~\bibnamefont {Mortessagne}}, \bibinfo
  {author} {\bibfnamefont {U.}~\bibnamefont {Kuhl}}, \bibinfo {author}
  {\bibfnamefont {S.}~\bibnamefont {Barkhofen}}, \bibinfo {author}
  {\bibfnamefont {S.}~\bibnamefont {Gehler}}, \bibinfo {author} {\bibfnamefont
  {H.-J.}\ \bibnamefont {St\"ockmann}}, \bibinfo {author} {\bibfnamefont
  {I.}~\bibnamefont {Foulger}}, \bibinfo {author} {\bibfnamefont
  {S.}~\bibnamefont {Gnutzmann}}, \ and\ \bibinfo {author} {\bibfnamefont
  {G.}~\bibnamefont {Tanner}},\ }\href {\doibase
  10.1103/PhysRevLett.114.110501} {\bibfield  {journal} {\bibinfo  {journal}
  {Phys. Rev. Lett.}\ }\textbf {\bibinfo {volume} {114}},\ \bibinfo {pages}
  {110501} (\bibinfo {year} {2015})}\BibitemShut {NoStop}%
\bibitem [{\citenamefont {Ambainis}\ \emph {et~al.}(2015)\citenamefont
  {Ambainis}, \citenamefont {Portugal},\ and\ \citenamefont
  {Nahimov}}]{ambainis2013search}%
  \BibitemOpen
  \bibfield  {author} {\bibinfo {author} {\bibfnamefont {A.}~\bibnamefont
  {Ambainis}}, \bibinfo {author} {\bibfnamefont {R.}~\bibnamefont {Portugal}},
  \ and\ \bibinfo {author} {\bibfnamefont {N.}~\bibnamefont {Nahimov}},\
  }\href@noop {} {\bibfield  {journal} {\bibinfo  {journal} {Quantum
  Information \& Computation}\ }\textbf {\bibinfo {volume} {15}},\ \bibinfo
  {pages} {1233} (\bibinfo {year} {2015})}\BibitemShut {NoStop}%
\bibitem [{\citenamefont {Portugal}\ \emph {et~al.}(2015)\citenamefont
  {Portugal}, \citenamefont {Boettcher},\ and\ \citenamefont
  {Falkner}}]{portugal2015one}%
  \BibitemOpen
  \bibfield  {author} {\bibinfo {author} {\bibfnamefont {R.}~\bibnamefont
  {Portugal}}, \bibinfo {author} {\bibfnamefont {S.}~\bibnamefont {Boettcher}},
  \ and\ \bibinfo {author} {\bibfnamefont {S.}~\bibnamefont {Falkner}},\ }\href
  {\doibase 10.1103/PhysRevA.91.052319} {\bibfield  {journal} {\bibinfo
  {journal} {Phys. Rev. A}\ }\textbf {\bibinfo {volume} {91}},\ \bibinfo
  {pages} {052319} (\bibinfo {year} {2015})}\BibitemShut {NoStop}%
\bibitem [{\citenamefont {Suzuki}(1976)}]{suzuki1976generalized}%
  \BibitemOpen
  \bibfield  {author} {\bibinfo {author} {\bibfnamefont {M.}~\bibnamefont
  {Suzuki}},\ }\href@noop {} {\bibfield  {journal} {\bibinfo  {journal}
  {Communications in Mathematical Physics}\ }\textbf {\bibinfo {volume} {51}},\
  \bibinfo {pages} {183} (\bibinfo {year} {1976})}\BibitemShut {NoStop}%
\bibitem [{\citenamefont {Suzuki}(1985)}]{suzuki1985decomposition}%
  \BibitemOpen
  \bibfield  {author} {\bibinfo {author} {\bibfnamefont {M.}~\bibnamefont
  {Suzuki}},\ }\href@noop {} {\bibfield  {journal} {\bibinfo  {journal}
  {Journal of mathematical physics}\ }\textbf {\bibinfo {volume} {26}},\
  \bibinfo {pages} {601} (\bibinfo {year} {1985})}\BibitemShut {NoStop}%
\bibitem [{\citenamefont {Lvovsky}\ and\ \citenamefont
  {Raymer}(2009)}]{lvovsky2009continuous}%
  \BibitemOpen
  \bibfield  {author} {\bibinfo {author} {\bibfnamefont {A.~I.}\ \bibnamefont
  {Lvovsky}}\ and\ \bibinfo {author} {\bibfnamefont {M.~G.}\ \bibnamefont
  {Raymer}},\ }\href@noop {} {\bibfield  {journal} {\bibinfo  {journal}
  {Reviews of Modern Physics}\ }\textbf {\bibinfo {volume} {81}},\ \bibinfo
  {pages} {299} (\bibinfo {year} {2009})}\BibitemShut {NoStop}%
\bibitem [{\citenamefont {Pollard}(2002)}]{pollard2002user}%
  \BibitemOpen
  \bibfield  {author} {\bibinfo {author} {\bibfnamefont {D.}~\bibnamefont
  {Pollard}},\ }\href@noop {} {\emph {\bibinfo {title} {A user's guide to
  measure theoretic probability}}},\ Vol.~\bibinfo {volume} {8}\ (\bibinfo
  {publisher} {Cambridge University Press},\ \bibinfo {year}
  {2002})\BibitemShut {NoStop}%
\bibitem [{\citenamefont {Suzuki}(1977)}]{suzuki1977convergence}%
  \BibitemOpen
  \bibfield  {author} {\bibinfo {author} {\bibfnamefont {M.}~\bibnamefont
  {Suzuki}},\ }\href@noop {} {\bibfield  {journal} {\bibinfo  {journal}
  {Communications in Mathematical Physics}\ }\textbf {\bibinfo {volume} {57}},\
  \bibinfo {pages} {193} (\bibinfo {year} {1977})}\BibitemShut {NoStop}%
\bibitem [{\citenamefont {Nielsen}\ and\ \citenamefont
  {Chuang}(2010)}]{nielsen2010quantum}%
  \BibitemOpen
  \bibfield  {author} {\bibinfo {author} {\bibfnamefont {M.~A.}\ \bibnamefont
  {Nielsen}}\ and\ \bibinfo {author} {\bibfnamefont {I.~L.}\ \bibnamefont
  {Chuang}},\ }\href@noop {} {\emph {\bibinfo {title} {Quantum computation and
  quantum information}}}\ (\bibinfo  {publisher} {Cambridge university press},\
  \bibinfo {address} {New York},\ \bibinfo {year} {2010})\BibitemShut {NoStop}%
\end{thebibliography}%

\end{document}